\documentclass[twocolumn,aps,prl,preprintnumbers]{revtex4}
\usepackage{}%
\usepackage{mathrsfs}
\usepackage{amsmath}
\usepackage{amsfonts}
\usepackage{amssymb}
\usepackage{harpoon}
\usepackage{amsthm}
\usepackage{graphicx}
\usepackage{bm}
\usepackage{color}%
\usepackage[colorlinks, citecolor=blue]{hyperref}
\providecommand{\U}[1]{\protect\rule{.1in}{.1in}}
\begin{document}
\title{Edge states in a two-dimensional honeycomb lattice of massive magnetic skyrmions}
\author{Z.-X. Li}
\author{C. Wang}
\author{Yunshan Cao}
\author{Peng Yan}
\email[Corresponding author: ]{yan@uestc.edu.cn}
\affiliation{School of Electronic Science and Engineering and State Key Laboratory of Electronic Thin Film and Integrated Devices, University of Electronic Science and Technology of China, Chengdu 610054, China}

\begin{abstract}
We study the collective dynamics of a two-dimensional honeycomb lattice of magnetic skyrmions. By performing large-scale micromagnetic simulations, we find multiple chiral and non-chiral edge modes of skyrmion oscillations in the lattice. The non-chiral edge states are due to the Tamm-Shockley mechanism, while the chiral ones are topologically protected against structure defects and hold different handednesses depending on the mode frequency. To interpret the emerging multiband nature of the chiral edge states, we generalize the massless Thiele's equation by including a second-order inertial term of skyrmion mass as well as a third-order non-Newtonian gyroscopic term, which allows us to model the band structure of skrymion oscillations. Theoretical results compare well with numerical simulations. Our findings uncover the importance of high order effects in strongly coupled skyrmions and are helpful for designing novel skyrmionic topological devices.
\end{abstract}

\maketitle
\emph{Introduction.$-$}In recent years, topological insulators have attracted considerably growing interest owing to their outstanding physical properties \cite{Hasan2010,Qi2011,Zhang2009,Hsieh2009,Pribiag2015}. The most peculiar character of topological insulators is that they can support chiral edge states which are absent in conventional insulators. Topological edge states are modes confined at the boundary of a system and generally have a certain chirality which enables them to be immune from small disturbances such as disorders and/or defects \cite{Hasan2010,Qi2011}. Ever since its discovery in electronic system \cite{Kane2005,Mele2005}, the topological edge state has been readily predicted and observed in optics \cite{Lu2014,Harari2018,Bandres2018}, mechanics \cite{Paulose2015,Nash2015,Mitchell2018}, acoustics \cite{Yang2015,He2016,Fleury2016} and very recently in magnetics \cite{Zhang2013,Mook2014,Kim2016,Owerre2016}.

There are two important excitations in magnetic systems. One is the spin wave (or magnon), the collective motion of magnetic moments. Magnon Hall effect and topology have been predicted in ferromagnetic insulators \cite{Zhang2013,Henk2014,Loss2017}. Shindou \emph{et al}. \cite{Shindou2013} and Wang \emph{et al}. \cite{Wang2017} demonstrated chiral spin-wave edge modes in a two-dimensional (honeycomb or square) lattice of spins coupled by magnetic dipole-dipole interactions. Chiral spin-wave edge states were also studied in the triangular skyrmion crystal \cite{Roldan2016}. The concept of magnonic topological insulator can be generalized to antiferromagnets as well \cite{Owerre2017,Nakata2017,smejkal2018}. The other one is the magnetic soliton, such as the magnetic vortex \cite{Wachowiak2002,Pribiag2007}, bubble \cite{Petit2015,Moon2015}, and skyrmion \cite{Muhlbauer2009,Jiang2015,Yang2018PRB,Yang2018PRL}. It has been shown that the collective gyration of magnetic solitons in an array resembles an one-dimensional (1D) wave \cite{Han2013,Yang2017,Kim2018}. In Ref. \cite{Kim2017}, Kim and Tserkovnyak generalized the situation to 2D: they theoretically studied the coupled oscillations of magnetic vortices and bubbles in a honeycomb lattice. By mapping the massless Thiele's equation into the Haldane model \cite{Haldane1988}, they predicted the chiral edge modes near the gyration frequency of the single soliton. However, it is well known that magnetic bubbles and skyrmions in particular manifest an inertia in their gyration motion \cite{Makhfudz2012,Yang2018OE}. The mass effect thus should be taken into account for developing a full theory on the coupled skyrmion oscillations. In addition, a direct comparison with a more rigorous micromagnetism approach is still lacking, in part due to the difficulty of simulating a very large spin system numerically.

In this work, we present both analytical and numerical studies of the collective dynamics of a 2D honeycomb lattice of magnetic skyrmions. From large-scale micromagnetic simulations, we identify multiple edge states below a cut-off frequency 20 GHz. Further, we find that one of them is trivially non-chiral and is due to the Tamm-Shockley mechanism, while the rest are chiral with different handednesses. To understand the multiband nature of the emerging chiral edge states, we generalize the Thiele's theory by including both a second-order inertial term of skyrmion mass and a third-order non-Newtonian gyroscopic term. We theoretically compute the band structure of the collective skyrmion gyrations. The results obtained agree very well with micromagnetic simulations.
\begin{figure}[ptbh]
\begin{centering}
\includegraphics[width=0.48\textwidth]{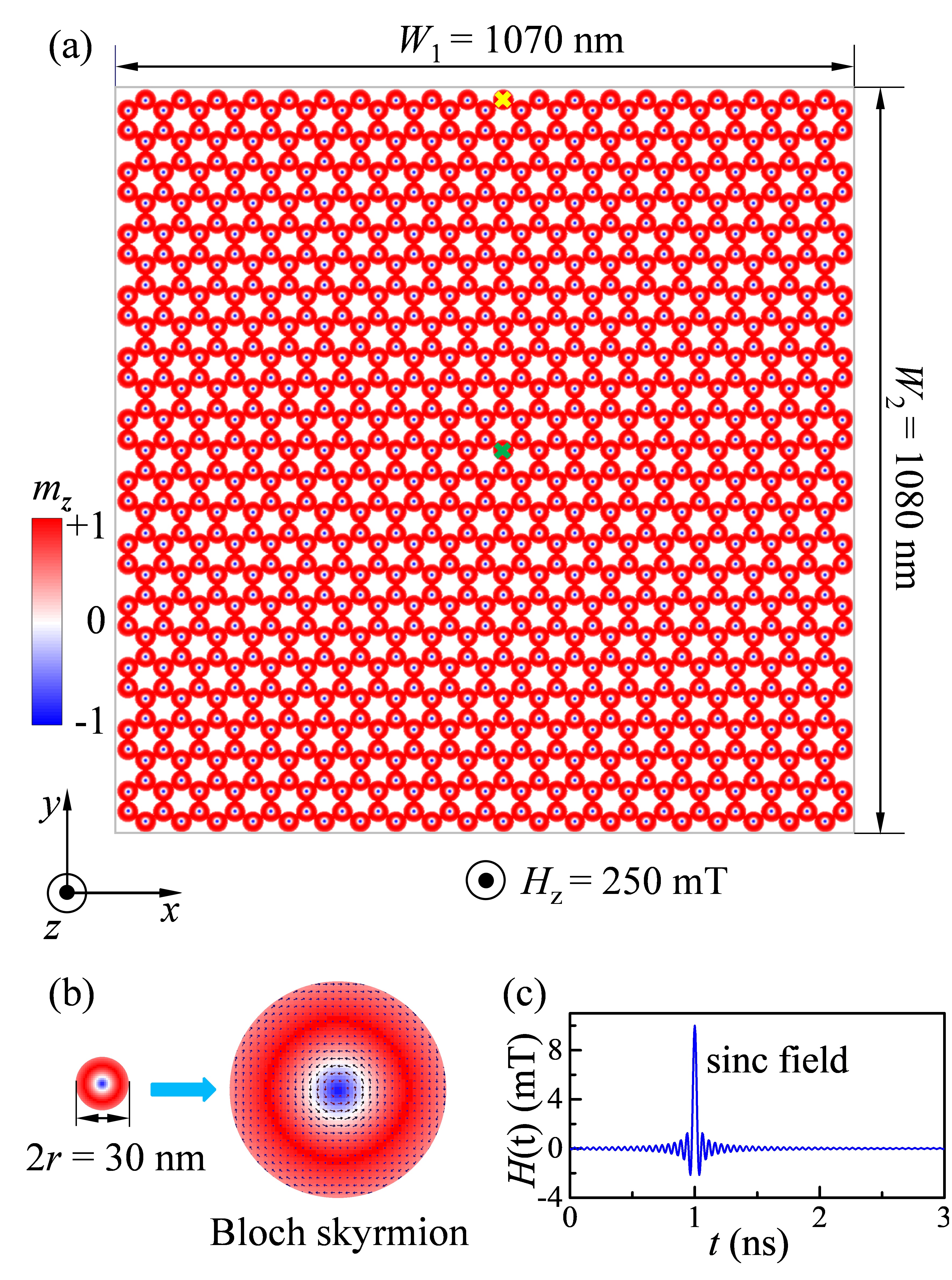}
\par\end{centering}
\caption{(a) Illustration of the honeycomb lattice with size $1070\times1080$ nm$^{2}$, including 984 Bloch skyrmions. A uniform magnetic field is applied along the $z$ axis to stabilize the skyrmions. Green and yellow crosses denote the positions of the driving fields in the center and at the edge of the lattice, respectively. (b) Zoomed in details of a nanodisk containing a Bloch skyrmion. (c) Time-dependence of the sinc-function field $H(t)$.}
\label{Figure1}
\end{figure}

\begin{figure}[ptbh]
\begin{centering}
\includegraphics[width=0.48\textwidth]{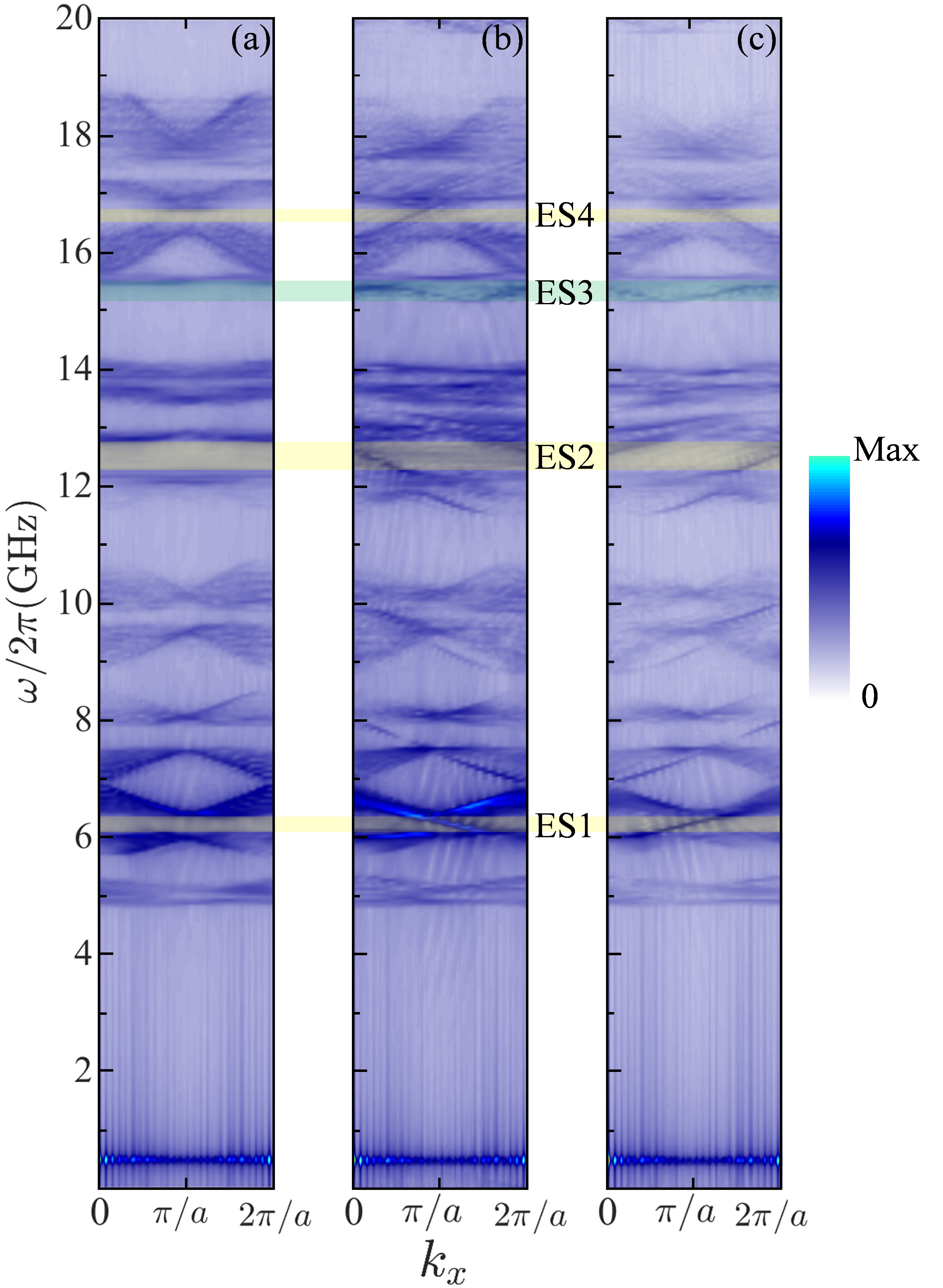}
\par\end{centering}
\caption{The band structure of skyrmion gyrations when the exciting field is in the film center (a) and at the film edge by evaluating the Fourier spectrum over the upper (b) and the lower (c) parts of the honeycomb lattice. The constant $a=2\sqrt{3}r$ represents the distance between the second-nearest neighboring nanodisks.}
\label{Figure2}
\end{figure}

\emph{Large-scale micromagnetic simulations.$-$}We consider a large 2D honeycomb lattice consisting of 984 identical magnetic nanodisks with the diameter of 30 nm and the thickness of 1 nm, as shown in Fig. \ref{Figure1}(a). Each disk contains a single magnetic skyrmion made of MnSi \cite{Tomasello2014} which supports the Bloch-type skyrmion [depicted in Fig. \ref{Figure1}(b)] due to the bulk Dzyakoshinskii-Moriya interaction (DMI) \cite{Yu2010,Seki2012}. The distance between nearest-neighbor disks is chosen to be equal to the disk diameter, indicating that nearest-neighbor skyrmions can strongly interact with each other mediated by the exchange spin-wave. It is worth noting that the dipolar interaction can not efficiently couple skyrmions when we artificially leave a physical gap/distance between nearest-neighboring nanodisks even as short as 1nm. What we observed is no more than the spectrum of isolated skyrmion gyrations, instead of collective ones (not shown). We performed micromagnetic simulations with MuMax3 \cite{Vansteenkiste2014}. The following materials parameters are used \cite{Tomasello2014}: the saturation magnetization $M_{s}=1.52\times10^{5}$ A/m, the exchange stiffness  $A=3.2\times10^{-13}$ J/m, the strength of the bulk DMI $D=0.15$ mJ/m$^{2}$, and the Gilbert damping constant $\alpha=0.01$ if not stated otherwise. To stabilize skyrmions in the nanodisks, we apply an external magnetic field perpendicular to the disk plane ($z$ axis), with the strength $H_{z}= 250$ mT. In the simulations, we set the cell size to be $1\times1\times1 $ nm$^{3}$. In order to excite the full spectrum (up to a cut-off frequency) of the skyrmion oscillations, we applied a sinc-function magnetic field $H(t)=H_{0}\sin[2\pi$\emph{f}$(t-t_{0})]/[2\pi$\emph{f}$(t-t_{0})]$ [see Fig. \ref{Figure1}(c)] along the $x$ direction with $H_{0}=10$ mT, $f=20$ GHz, and $t_{0}=1$ ns. The exciting field is applied locally on a disk which is either in the center or at the edge of the lattice, as shown in Fig. \ref{Figure1}(a) by the green and yellow crosses, respectively. The spatiotemporal evolutions of the skyrmion guiding centers $\textbf{R}_{j}=(R_{j,x}, R_{j,y}$) in all nanodisks are recorded every 5 ps. Here $R_{j,x}$ and $R_{j,y}$ are defined by  $R_{j,x}=\frac {\int \!\!\! \int{xqdxdy}}{\int \!\!\! \int{qdxdy}}$ and $R_{j,y}=\frac {\int \!\!\! \int{yqdxdy}}{\int \!\!\! \int{qdxdy}}$, with $q=\textbf{m}\cdot(\frac {\partial \textbf{m}}{\partial {x} } \times \frac {\partial \textbf{m}}{\partial y } )$ being the skyrmion charge density \cite{Makhfudz2012,Dai2014} and $\textbf{m}$ the unit vector of the local magnetic moment. The integral region is confined in the $j$-th nanodisk.

To obtain the dispersion relation of skyrmion gyrations, we compute the spatiotemporal Fourier spectrum of the skyrmion positions over the lattice. Figure \ref{Figure2}(a) shows the simulated band structure of skyrmion oscillation when the exciting field locates in the lattice center. We find that there is no bulk state in the gaps (areas shaded in both yellow and green). We then put the driving field at the edge of lattice. Implementing the Fourier analysis over the upper ($W_{2}/2<y<W_{2}$) and the lower parts ($0<y<W_{2}/2$) of the lattice, with results plotted in Fig. \ref{Figure2}(b) and Fig. \ref{Figure2}(c), respectively, we find it interesting that four edge states emerge in the spectrum gaps, labeled as ES1-ES4. By evaluating the group velocity  $d\omega/dk_{x}$ of each mode with $\omega$ the frequency and $k_{x}$ the wave vector along $x$ direction, we conclude that three edge states ES1, ES2 and ES4 (shaded in yellow) are unidirectional and chiral, in which ES1 and ES2 counterclockwise propagate, while ES4 behaves oppositely. However, ES3 (shaded in green) is bidirectional and thus non-chiral. Further, we judge that the edge modes near $7.7$ GHz and $9.8$ GHz shown in Figs. \ref{Figure2}(b) and (c) are induced by higher-order effects [see Fig. \ref{Figure4}(a) and discussions below].

\begin{figure}[ptbh]
\begin{centering}
\includegraphics[width=0.48\textwidth]{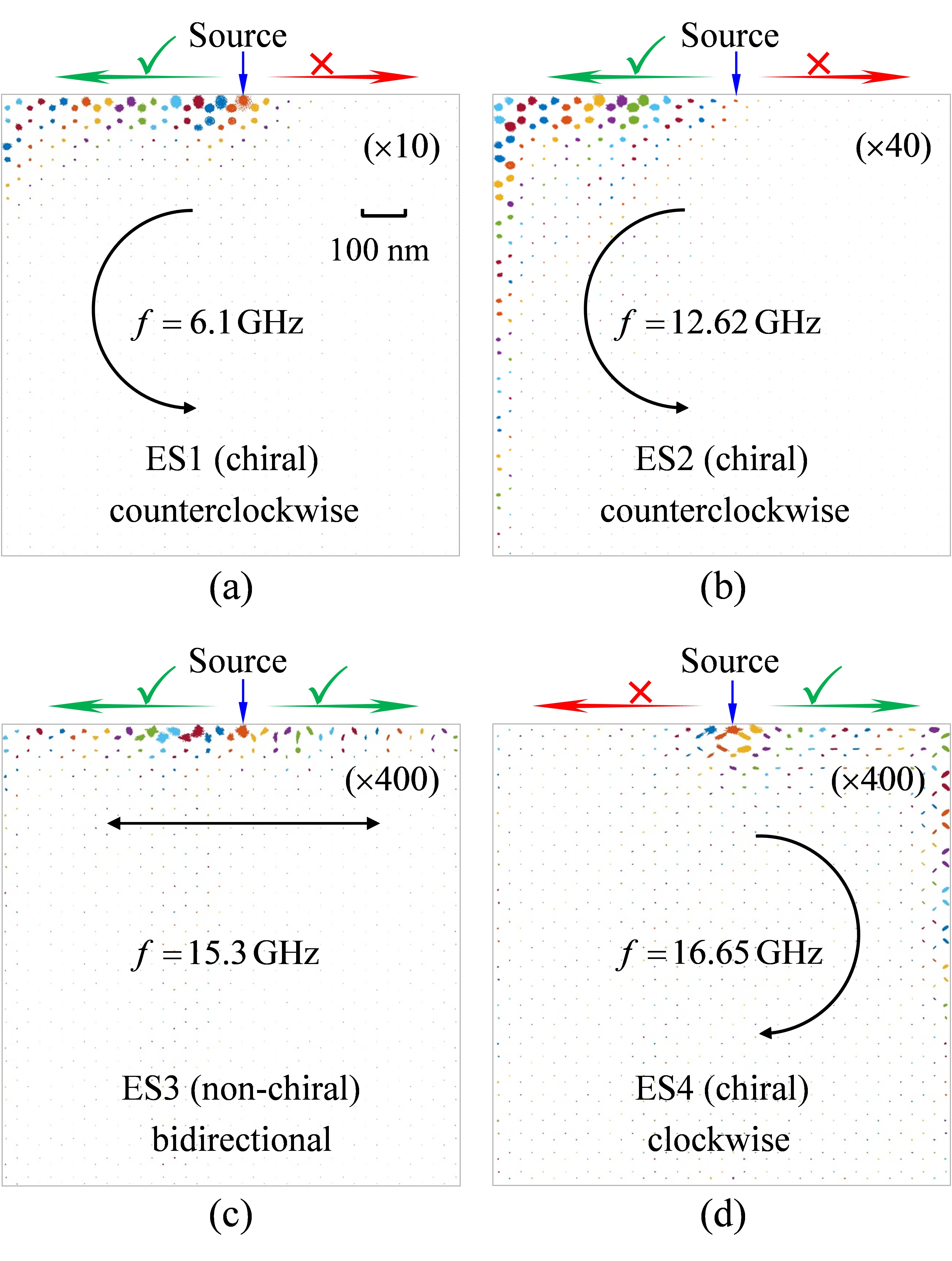}
\par\end{centering}
\caption{Snapshot of the propagation of edge states with frequency $\emph{f}=6.1$ GHz (a), $12.62$ GHz (b), $15.3$ GHz (c), and $16.65$ GHz (d) at $t=40$ ns. Since the oscillation amplitudes of the skyrmion guiding centers are too small, we have magnified them by 10, 40 or 400 times labeled in each figure, correspondingly.}
\label{Figure3}
\end{figure}
To visualize the propagation of the edge states, we choose four representative frequencies: 6.1 GHz for ES1, 12.62 GHz for ES2, 15.3 GHz for ES3, and 16.65 GHz for ES4, and simulate their dynamics by a sinusoidal field $\textbf{h}(t)=h_0\sin(2\pi ft)\hat{x}$ with $h_0=10$ mT applied on the disk at the edge, indicated by the blue arrows in Fig. \ref{Figure3}. Figures \ref{Figure3}(a), (b) and (d) show the propagation of chiral edge states. One can clearly observe unidirectional wave propagation of these modes with either a counterclockwise manner [ES1 and ES2 shown in Figs. \ref{Figure3}(a) and (b), respectively] or a clockwise one [ES4 plotted in Fig. \ref{Figure3}(d)]. It should be noted that we have adopted artificially a rather small Gilbert damping constant $\alpha=10^{-5}$ when simulating the $\emph{f}$ = 6.1 GHz mode in Fig. \ref{Figure3}(a), since the decay length of the mode is too short to show its chirality clearly at $\alpha=0.01$. We thus report the coexistence of multiband edge states with opposite chiralities in a given soliton lattice, to the best of our knowledge, for the first time. An analytical understanding of these numerical results shall be given below. In contrast, the propagation of ES3 is bidirectional, as shown in Fig. \ref{Figure3}(c). This non-chiral mode can be simply explained in terms of the Tamm-Shockley mechanism \cite{Tamm1932,Shockley1939} which predicts that the periodicity breaking of the crystal potential at the boundary can lead to the formation of a conducting surface/edge state. Further, we confirmed that the propagation of the chiral mode is immune from the artificial lattice defects and robust against the type of boundary, while the Tamm-Shockley mode is not (see Supplemental Material Sec. A for details \cite{Supp}).

\emph{Theoretical model.$-$}To theoretically understand the multiband chiral skyrmionic edge states carrying opposite chiralities, we need to generalize the original massless Thiele's equation \cite{Thiele1973}. To this end, we assume that the steady-state magnetization $\textbf{m}$ depends on not only the position of the guiding center $\textbf{R}(t)$ but also its velocity $\dot{\textbf{R}}(t)$ and acceleration $\ddot{\textbf{R}}(t)$, and write $\textbf{m}=\textbf{m}(\textbf{r}-\textbf{R}(t),\dot{\textbf{R}}(t),\ddot{\textbf{R}}(t))$. After some algebra, we obtain the corrections of both a second-order inertial term of skyrmion mass $M$ \cite{Makhfudz2012,Yang2018OE,Buttner2015} and a non-Newtonian third-order gyroscopic term $G_{3}$ \cite{Mertens1997,Ivanov2010,Cherepov2012}  (see Supplemental Material Sec. B for detailed derivations \cite{Supp}):
\begin{equation}\label{Eq1}
  G_{3}\hat{z}\times\frac{d^{3}\textbf{U}_{j}}{dt^{3}}-M\frac{d^{2}\textbf{U}_{j}}{dt^{2}}+G\hat{z}\times \frac{d\textbf{U}_{j}}{dt}+\textbf{F}_{j}=0,
\end{equation}
where $\mathbf{U}_{j}\equiv \mathbf R_{j} - \mathbf R_{j}^{0}$ is the displacement of the skyrmion from its equilibrium position $\mathbf R_{j}^{0}$ and $G = -4\pi$$Qd M_{s}$/$\gamma$ is the gyroscopic constant with $Q=\int \!\!\! \int{qdxdy}$ being the skyrmion charge [$Q=+1$ for the skyrmion configuration shown in Fig. \ref{Figure1}(b)], $d$ the thickness of nanodisk, and $\gamma$ the gyromagnetic ratio. For MnSi \cite{Tomasello2014}, we have $G=1.091\times10^{-14}$ Js/m$^{2}$. The conservative force is expressed as $\textbf{F}_{j}=-\partial W / \partial \mathbf U_{j}$ where $W$ is the potential energy as a function of the displacement: $W=\sum_{j}K\textbf{U}_{j}^{2}/2+\sum_{j\neq k}U_{jk}/2$ with $U_{jk}=I_{\parallel}U_{j}^{\parallel}U_{k}^{\parallel}-I_{\perp}U_{j}^{\perp}U_{k}^{\perp}$ \cite{Kim2017,Shibata2003,Shibata2004}. Here $I_{\parallel}$ and $I_{\perp}$ are the longitudinal and the transverse coupling constants, respectively.

We therefore obtain the equation of motion for $\mathbf{U}_{j}\equiv(u_{j},v_{j})$:
\begin{equation}\label{Eq2}
 \begin{aligned}
  &G_{3}\left(
          \begin{array}{c}
            -\dddot{v}_{j} \\
            \dddot{u}_{j} \\
          \end{array}
        \right)-M\left(
          \begin{array}{c}
            \ddot{u}_{j} \\
            \ddot{v}_{j} \\
          \end{array}
        \right)+G\left(
          \begin{array}{c}
            -\dot{v}_{j} \\
            \dot{u}_{j} \\
          \end{array}
        \right)=K\left(\begin{array}{c}
                   u_{j} \\
                   v_{j}
                 \end{array}
                 \right)\\
 &+\sum_{k\in\langle j\rangle}\left(
                                                                                                                        \begin{array}{cc}
                                                                                                                          \zeta+\xi\cos2\theta_{jk} & \xi\sin2\theta_{jk} \\
                                                                                                                          \xi\sin2\theta_{jk} & \zeta-\xi\cos2\theta_{jk}\\
                                                                                                                        \end{array}
                                                                                                                      \right)\left(
                                                                                                                               \begin{array}{c}
                                                                                                                                 u_{k} \\
                                                                                                                                 v_{k} \\
                                                                                                                               \end{array}
                                                                                                                             \right),\\
                                                                                                                              \end{aligned}
\end{equation}
where $\zeta=(I_{\parallel}-I_{\perp})/G$, $\xi=(I_{\parallel}+I_{\perp})/G$, $\theta_{jk}$ is the angle of the direction $\hat{e}_{jk}$ from $x$ axis with $\hat{e}_{jk}=(\mathbf{R}_{k}^{0}-\mathbf{R}_{j}^{0})/|\mathbf{R}_{k}^{0}-\mathbf{R}_{j}^{0}|$, and $\langle j\rangle$ is the set of the nearest neighbors of $j$. By defining $\psi_{j}=u_{j}+i v_{j}$, we have
\begin{equation}\label{Eq3}
   \hat{\mathcal {D}}\psi_{j}=\omega_{K}\psi_{j}+\sum_{k\in\langle j\rangle}(\zeta\psi_{k}+\xi e^{i2\theta_{jk}}\psi^{*}_{k}),
\end{equation}
with the differential operator $\hat{\mathcal {D}}=i\omega_{3}\frac{d^{3}}{dt^{3}}-\omega_{M}\frac{d^{2}}{dt^{2}}+i\frac{d}{dt}$, $\omega_{3}=G_{3}/G$, $\omega_{M}=M/G $, and $\omega_{K}=K/G$. We then expand the complex variable to
\begin{equation}\label{Eq4}
  \psi_{j}=\chi_{j}(t)\exp(-i\omega_{0}t)+\eta_{j}(t)\exp(i\omega_{0}t).
\end{equation}
For counterclockwise (clockwise) skyrmion gyrations, one can justify $|\chi_{j}|\gg|\eta_{j}|$ ($|\chi_{j}|\ll|\eta_{j}|$). Substituting Eq. \eqref{Eq4} into Eq. \eqref{Eq3}, we obtain the following  eigenvalue equation
\begin{equation}\label{Eq5}
  \begin{aligned}
 \hat{\mathcal {D}}\psi_{j}=&(\omega_{K}-3\xi^{2}/2\bar{\omega}_{K})\psi_{j}+\zeta\sum_{k\in\langle j\rangle}\psi_{k}\\
  &-(\xi^{2}/2\bar{\omega}_{K})\sum_{l\in\langle\langle j\rangle\rangle}\cos(2\bar{\theta}_{jl})\psi_{l}\\
  &-i(\xi^{2}/2\bar{\omega}_{K})\sum_{l\in\langle\langle j\rangle\rangle}\sin(2\bar{\theta}_{jl})\psi_{l},
  \end{aligned}
\end{equation}
imposing $\omega_{0}$ to satisfy the following condition
\begin{equation}\label{Eq6}
\omega_{K}=\omega_{M}\omega^{2}_{0}\pm(\omega_{3}\omega^{3}_{0}-\omega_{0}),
\end{equation}
for clockwise and counterclockwise skyrmion gyrations, respectively, with $\bar{\omega}_{K}=\omega_{K}-\omega^{2}_{0}\omega_{M}$, $\bar{\theta}_{jl}=\theta_{jk}-\theta_{kl}$ the relative angle from the bond $k\rightarrow l$ to the bond $j\rightarrow k$ with $k$ between $j$ and $l$, and $\langle\langle j\rangle\rangle$ the set of the second-nearest neighbors of $j$.
\begin{figure}[ptbh]
\begin{centering}
\includegraphics[width=0.48\textwidth]{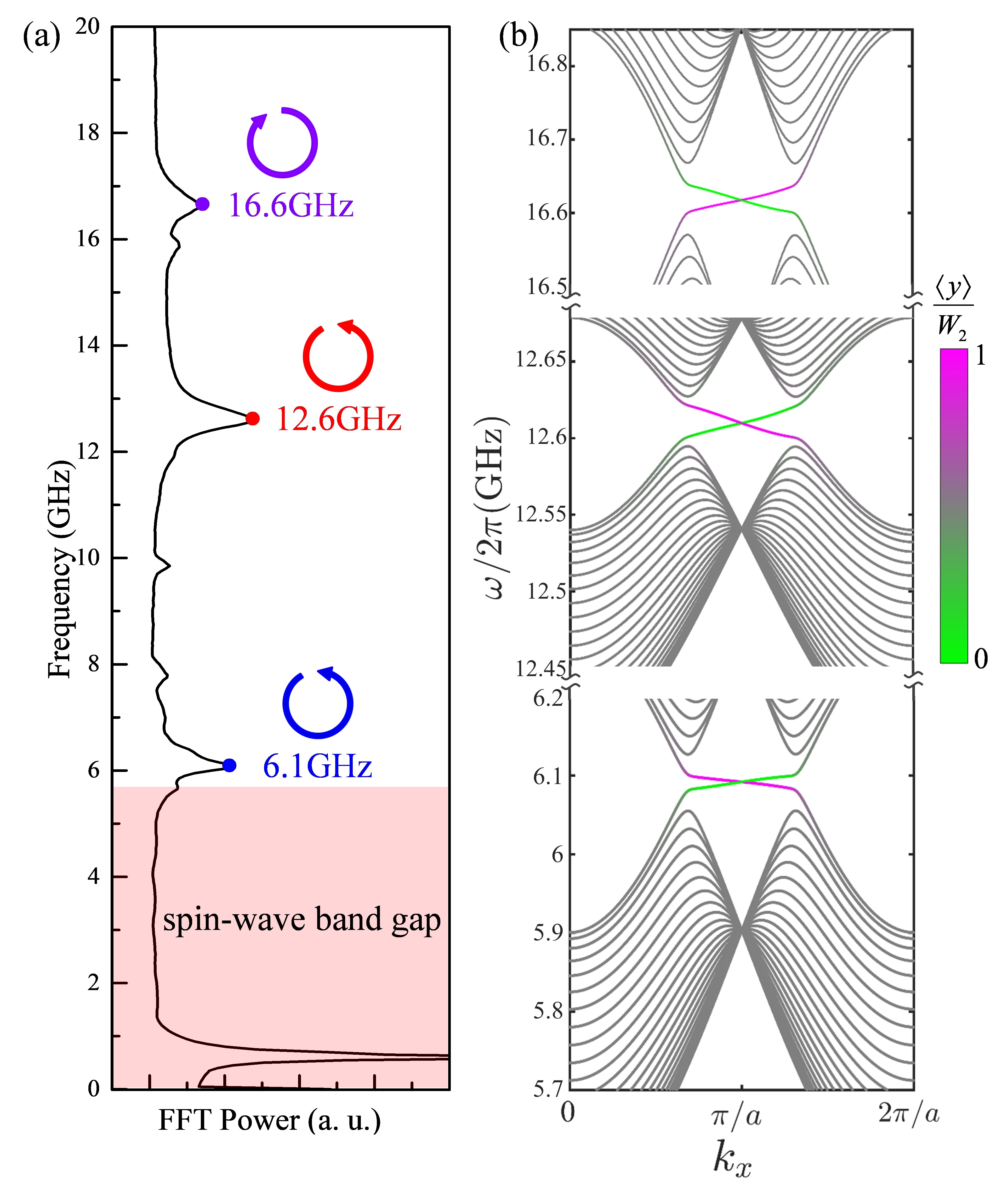}
\par\end{centering}
\caption{(a) Resonant spectrum of skyrmion gyrations when the exciting field is applied over the whole system. Inset shows the chirality/handedness of the skyrmion guiding center of each mode. (b) Band structure by solving Eq. \eqref{Eq5}.}
\label{Figure4}
\end{figure}

Our task then is to determine the three key parameters $G_{3}$, $M$ and $K$, which reflect the global resonances properties of the whole lattice. To this end, we apply the sinc-function field over the entire system to excite the collective skyrmion oscillations. Numerical simulations show three strong resonance peaks at $\omega_{0,1}/2\pi=6.1$ GHz, $\omega_{0,2}/2\pi=12.6 $ GHz, and $\omega_{0,3}/2\pi=16.6$ GHz above the spin-wave band gap ($\sim5.8$ GHz caused by both the applied static field and the demagnetization field), as plotted in Fig. \ref{Figure4}(a). The chirality of skyrmion gyrations in each mode is marked in the figure. Two weak peaks close to $7.7$ GHz and $9.8$ GHz are due to higher order effects and are not the scope of the present work. We also note another low-frequency mode at $0.6$ GHz localized in the spin-wave gap, which corresponds to the gyration of an isolated skyrmion and is irrelevant to the collective edge state. We can thus obtain the inertial mass $M=1.5612\times10^{-26}$ kg, the spring constant $K =3.7446\times10^{-4}$J/m$^{2}$, and the third-order gyroscopic parameter $G_{3}=1.1832\times10^{-36}$Js$^{3}$/m$^{2}$ by solving Eq. \eqref{Eq6}. Further, by fitting the interaction energy of two skyrmion disks with the formula of $W$, we self-consistently obtain the other two important parameters, $I_{\parallel}=-1.965\times10^{-6}$J/m$^{2}$ and $I_{\perp}=-1.191\times10^{-5}$J/m$^{2}$ (see Supplemental Material Sec. C for calculations \cite{Supp}). Figure \ref{Figure4}(b) shows the computed band structure of the skyrmion gyrations near the resonance frequencies $\omega_{0}=\omega_{0,1}$, $\omega_{0,2}$, and $\omega_{0,3}$ by solving Eq. \eqref{Eq5} with the periodic boundary condition along $x$ direction and the zigzag termination at $y=0$ and $y=W_{2}$. We also calculate the average vertical position of the modes $\langle{y}\rangle\equiv\sum_{j}R^{0}_{j,y}|\textbf{U}_{j}|^2/\sum_{j}|\textbf{U}_{j}|^2$ where $R^{0}_{j,y}$ is the equilibrium position of the skyrmion projected onto the $y$ axis, represented by different colors: closer to magenta indicating more localized at the upper edge. It is interesting to note that the chirality of ES4 is opposite to those of ES1 and ES2, which is fully consistent with the results of micromagnetic simulations plotted in Figs. \ref{Figure2} and \ref{Figure3}. An interpretation is as following: The direction reversal of the skyrmion gyration generates a $\pi$ phase accumulation in the next-nearest-neighbor hopping term of Eq. \eqref{Eq5}. The Chern numbers of two neighbouring bulk bands then switch their signs, so that the chirality of the edge state in between reverses (see Supplemental Material Sec. D for details \cite{Supp}).

\emph{Conclusion.$-$}To conclude, we have studied the edge-state excitations in two-dimensional honeycomb lattices of magnetic skyrmions. By implementing large-scale micromagnetic simulations, we uncovered multiple edge states in the lattice, of which one mode is trivial and non-chiral while the rest are chiral and robust against lattice defects, and can carry opposite chiralities. To explain the multiband feature of these chiral edge modes, we generalized the Thiele's equation by considering both a mass term and a non-Newtonian gyroscopic one. Theoretical results compare very well with micromagnetic simulations. Our findings reveal the critical role of high order effects in strongly coupled skyrmions. The emerging multiband chiral edge modes possessing different handednesses should be appealing for designing future skyrmionic topological devices. We envision the existence of chiral edge modes in a skyrmion lattice of triangular type, which is an interesting issue for future study.

\begin{acknowledgments}
This work was supported by the National Natural Science Foundation of China (Grants No. 11604041 and 11704060), the National Key Research Development Program under Contract No. 2016YFA0300801, and the National Thousand-Young-Talent Program of China. C. W. is supported the China Postdoctoral Science Foundation (Grants No. 2017M610595 and 2017T100684) and the National Nature Science Foundation of China under Grant No. 11704061.
\end{acknowledgments}

\end{document}